\documentclass[10pt,twocolumn]{article}

\usepackage{multicol}
\usepackage{graphicx}
\usepackage{amsmath}
\usepackage{amssymb}
\usepackage{bm}
\usepackage{lipsum}
\usepackage{color}
\usepackage{siunitx}
\usepackage{authblk}
\usepackage[margin=2cm]{geometry}

\title{Homogeneous deposition of particles by absorption on hydrogels}
\author[1]{Fran\c{c}ois Boulogne}
\author[1]{Fran\c{c}ois Ingremeau}
\author[2]{Julien Dervaux}
\author[2]{Laurent Limat}
\author[1]{Howard A. Stone}
\affil[1]{Department of Mechanical and Aerospace Engineering, Princeton University, Princeton, NJ 08544}
\affil[2]{Laboratoire Mati\`ere et Syst\`emes Complexes (MSC), UMR 7057 CNRS, Universit\'e Paris Diderot, B\^atiment Condorcet, 10 rue Alice Domon et L\'eonie Duquet, Paris, France}
\date{\small\today}
\begin{document}


\twocolumn[
    \begin{@twocolumnfalse}
        \maketitle
        \begin{abstract}
    When a drop containing colloidal particles evaporates on a surface, a circular stain made of these particles is often observed due to an internal flow toward the contact line.
    To hinder this effect, several approaches have been proposed such as flow modification by addition of surfactants or control of the interactions between the particles.
    All of these strategies involve the liquid phase while maintaining the drying process.
    However, substitution of evaporation by absorption into the substrate of the solvent has been investigated less.
    Here, we show that a droplet containing colloidal particles deposited on swelling hydrogels can lead to a nearly uniform coating.
    We report experiments and theory to explore the relation between the gel swelling, uniformity of deposition and the adsorption dynamics of the particles at the substrate.
    Our findings suggest that draining the solvent by absorption provides a robust route to homogeneous coatings.
    ~\\~\\
        \end{abstract}
    \end{@twocolumnfalse}
]

It is well known that the accumulation of particles at a contact line can be explained by liquid flow in a droplet \cite{Deegan1997,Deegan2000a,Marin2011}, which rationalizes formation of a so called coffee stain.
    The combination of an anchored contact line and the liquid evaporation induces a radial flow, which is enhanced by the larger evaporation rate at the corner.
    To suppress the coffee stain effect, different strategies have been proposed \cite{Kuang2014}.
    For instance, a surfactant initially added in the liquid phase \cite{Kajiya2009,Still2012} or produced by bacteria \cite{Sempels2013} can reverse the internal flow field through Marangoni stresses related to surface tension gradients.
    Moreover, the shape of the colloidal particles can change the topology of the menisci at the liquid-air interface, which modifies the particle-particle interactions and suppresses the coffee stain effect for mixtures of spheres and ellipsoids \cite{Yunker2011}.
    Also, carefully chosen electric fields can be applied to manipulate the motion of the contact line and to apply an electrophoretic force on particles to suppress their accumulation at the contact line \cite{Wray2014}.

    \begin{figure}
        \centering
        \includegraphics[width=6cm]{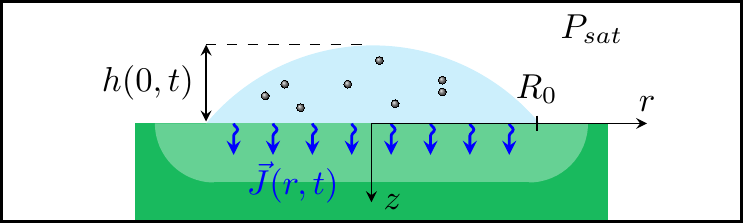}
        \caption{
            Schematic of the liquid absorption by a gel.
            The gel in its initial state is shown in dark green and the swollen gel in light green.
        }\label{fig:setup}
    \end{figure}

    At present, methods for suppressing the coffee stain effect mainly focus on the properties of the liquid phase, which evaporates.
    In contrast, our approach consists in removing the liquid with absorption by the substrate, which leaves the particles at the surface of the substrate.
    This strategy can be achieved with different materials such as solid porous networks \cite{Biot1941,Aradian2000,Bacri2000,Dou2012,Pack2015} for which the absorption is driven by capillarity.
 	However, the study of the swelling of gels involving a difference of chemical potential \cite{Doi2009} for particle deposition has been less investigated.
    Here we report results documenting absorption by hydrogels, where the adsorption kinetics can be adjusted by changing the initial composition of the gel.
    Moreover, the small pore size of the gel prevents the particles from migrating into the material, which ensures spatially homogeneous absorbing properties and spatially homogeneous deposition patterns.

    Whereas the description of the contact line dynamics of a pure water droplet on hydrogels has received recent attention \cite{Kajiya2011}, the final particle distribution and the associated mechanism has not been investigated to date.
    Nevertheless, high quality coatings of gels are desirable in many emerging technologies and medical applications such as controlled-release drug delivery \cite{Hoare2008}, tissue engineering and regenerative medicine \cite{Hunt2014}.
In particular, colloidosomes consist of gel microspheres covered with small particles.
For molecules contained in these microspheres of gels, it has been shown that the diffusivity of these molecules through the gel interface is slowed down by the presence of particles \cite{Dinsmore2002}.

    \paragraph{Experimental protocol}

    In our experiments, the absorbing substrate consists of a polydimethyl acrylamide (PDMA) hydrogel.
    We denote $\Gamma_a$ the weight fraction of dimethyl acrylamide, which controls the swelling rate.
        Typically, we use a $3$ mm thick layer of gel, which is larger than the drop height ($h_0\sim 0.5$ mm).
    The colloidal suspension consists of orange fluorescent (540/560) polystyrene beads (FluoSpheres, Life Technologies, USA) of $2a=1.1$ $\mu$m diameter and with a density 1.05 g/cm$^3$.
    The suspension is diluted in pure water at a concentration $C_p$ in the range 2000 to 8000 particles/mm$^3$.
    The corresponding volume fractions are within the range $1 \times 10^{-6}$ to $4 \times 10^{-6}$.
    These very low values ensure that particle-particle interactions are negligible in the suspension.
    The P\'eclet number is defined as $\textrm{Pe} = a {\cal V}/D_0$ where the sedimentation speed is ${\cal V} = \Delta \rho g a^2 / \eta_s$, $\Delta \rho$ is the density difference, $\eta_s$ is the viscosity of the solvent and $D_0$ is the Stoke-Einstein diffusion coefficient\cite{Benes2007}. Thus, $\textrm{Pe}\approx 0.2$ and the sedimentation can be neglected.

\begin{figure}
        \centering
        \includegraphics[width=6.5cm]{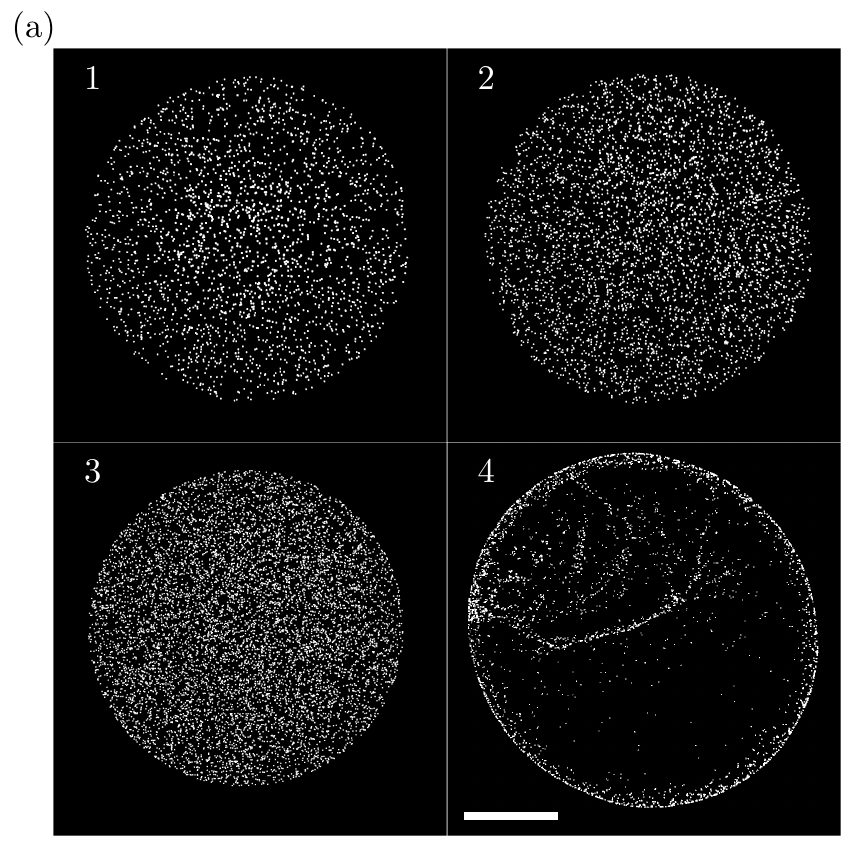}\\
        \includegraphics[width=7cm]{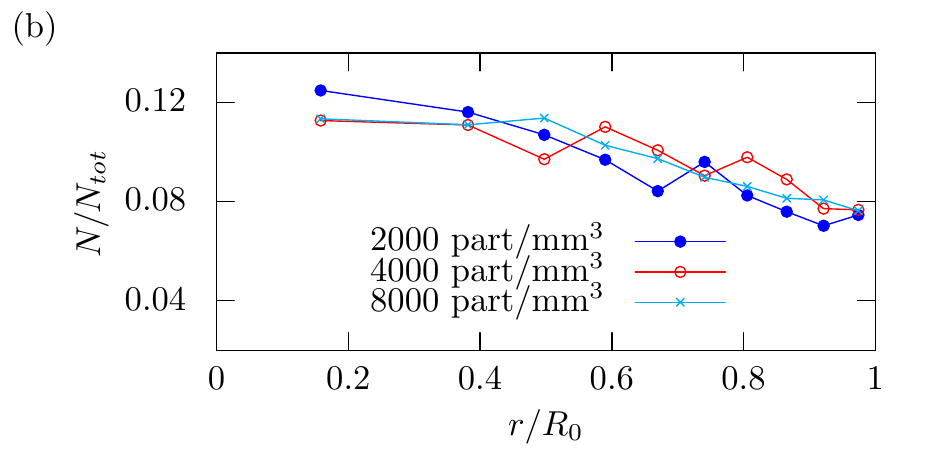}\\
        \caption{
            Deposition of particles on an absorbing hydrogel.
            (a) Images 1-3: final patterns on gels ($\Gamma_a = 0.15$) for various initial particle concentrations; $C_p= 2000,\,4000,\,8000$ particles/mm$^3$, respectively.
            Image 4: 1 $\mu$m diameter particles deposited on a cover glass slide from a drop of $0.8$ $\mu\ell$ at $C_p=4000$ particles/mm$^3$.
        The picture shows a coffee stain ring as well as a non-uniform deposit in the center due to stick-slip motion of the contact line.
        Scale bar $500$ $\mu$m.
            (b) Particle distribution $N/N_{tot}$ along the dimensionless radial position $r/R_0$ with $R_0$ the initial radius of the drop; $\Gamma_a=0.15$.
            The particle distribution is calculated for 10 concentric sectors of the same area and normalized by the total number of particles $N_{tot}$.
            Values are centered on each sectors.
        }\label{fig:final}
    \end{figure}

        A $0.8$ mm$^3$ colloidal drop is deposited on the surface of a hydrogel and the system is placed in an atmosphere saturated in water vapor preventing evaporation (Fig. \ref{fig:setup}).
    Once the drop is fully absorbed by the gel, the fluorescent particles are imaged by fluorescence microscopy.
    Images 1-3 of Fig. \ref{fig:final}a show the particles adsorbed at the surface of the gel for different particle concentrations. 
    Contrary to the coffee ring pattern usually observed on hard substrates \cite{Deegan1997} such as glass (Fig. \ref{fig:final}a, image 4),
    the final pattern resulting from the solvent absorption by the hydrogel is homogeneous.

    To analyze this observation in a quantitative way, we plot the particle distribution along the radial direction as shown in Fig. \ref{fig:final}b.
    The final particle distribution decreases weakly from the center to the edge of the drop (Fig. \ref{fig:final}b).
    Compared to the pattern obtained after evaporation of the same solution on a glass substrate (Fig. \ref{fig:final}a, picture 4), we can characterize the particle deposition on a hydrogel as nearly uniform.

    \paragraph{General dynamics and particle deposition}

\begin{figure}
        \centering
        \includegraphics[width=.95\linewidth]{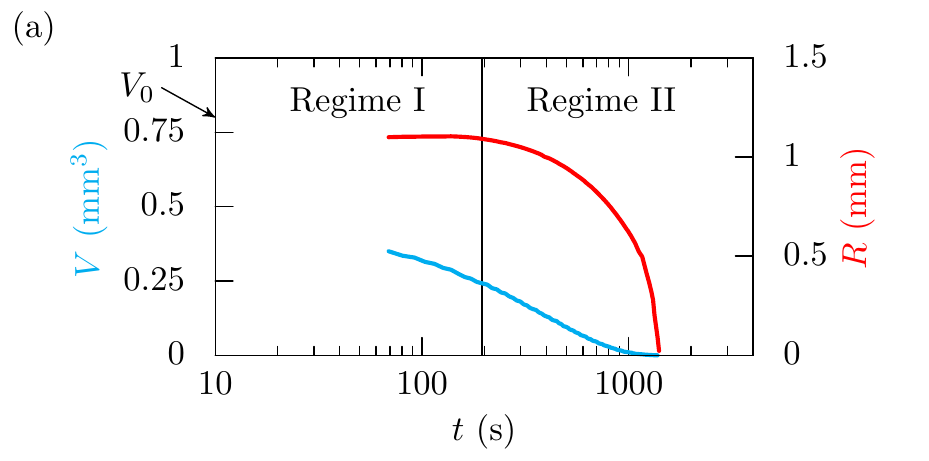}\\
        \includegraphics[height=5cm]{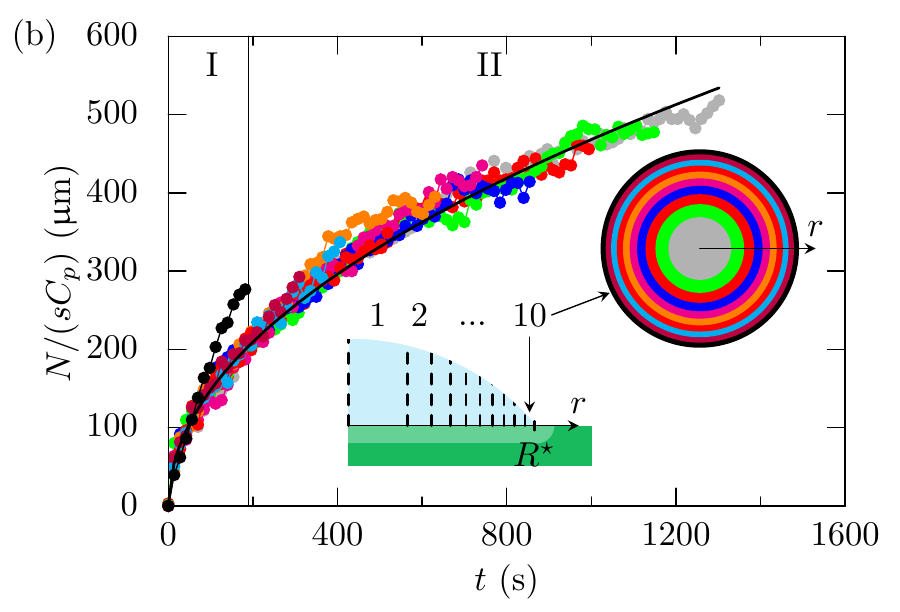}
        \caption{
            Dynamics of the liquid absorption.
            (a) Time evolution of the volume and radius of a $V_0=0.8$ mm$^3$ drop as measured using confocal microscopy.
            During the first $190$ s, the contact line is pinned and $70 \%$ of the volume is absorbed during this Regime $\textrm{I}$.
            After that time, the contact line recedes during Regime \textrm{II}.
            (b) Time evolution of the number of particles $N$ adsorbed on concentric rings of area $s = \pi R_0^{2}/10$ as measured using fluorescence microscopy.
            Regime I corresponds to the constant radius period whereas Regime II corresponds to the receding contact line period.
            Particle counting stops when the contact line enters in the corresponding ring.
            The deviation of the black points from the black solid line ($N = s C_p \sqrt{\kappa t}$) shows a higher flux at the drop edge.
            For both plots, $\Gamma_a = 0.20$ and $C_p=4000$ particles/mm$^3$.
        }\label{fig:kinetics}
    \end{figure}

    In order to understand the mechanism that leads to this pattern, we combine experimental and theoretical investigations to analyze the dynamics of liquid absorption in the gel in the absence of evaporation.
    The absorption of a drop deposited on a swelling hydrogel is composed of two successive regimes (Fig. \ref{fig:kinetics}a).
    First, the drop has a constant contact radius $R_0$ and the drop volume decreases in time because of absorption by the substrate, which tends to slightly deform the gel interface (Regime I).
    When the slope of the liquid-air interface matches that of the swollen hydrogel, the contact line starts to recede \cite{Kajiya2011} until complete absorption of the liquid occurs (Regime II).
    In our experimental conditions, the transition between Regimes I and II occurs for an absorbed volume that corresponds to $70$\% of the initial drop volume.

    We observed that particles are irreversibly adsorbed on the gel surface.
    First, adsorbed particles have no motion on the gel and second, these particles are not removed by the receding contact line during Regime II despite of the drag force applied on them \cite{GomezSuarez1999}.
    Also, particles do not penetrate in the hydrogel because the mesh size of the gel is much smaller than the particle size \cite{Sudre2011}.

    To investigate the final deposition pattern of particles, the local flux of water from the drop to the gel must be described.
    First, the main difference with the classic evaporation problem \cite{Deegan1997} concerns the diffusion process.
    As we will argue, the absorption timescale is shorter than the diffusion timescale, so that the diffusion problem is time dependent.
    Also, we observe that during the absorption of the liquid, particles in the bulk of the drop have an outer radial motion.
    This can be attributed to the pinned contact line, which imposes a radial flux of the solvent.
    Particles are Brownian and initially, the particle concentration $C_p$ is homogeneous in the drop.
    Because particles are irreversibly adsorbed on the surface of the gel, the number of particles deposited on the surface is proportional to the volume of solvent absorbed by the gel.
    Consequently, the particle concentration in the liquid remains constant in time.

    \begin{figure*}
        \centering
        \begin{minipage}[b]{0.45\linewidth}
        \includegraphics[width=\linewidth]{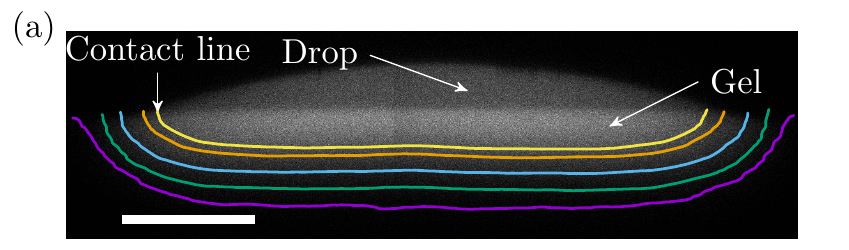}\\
        \includegraphics[width=\linewidth]{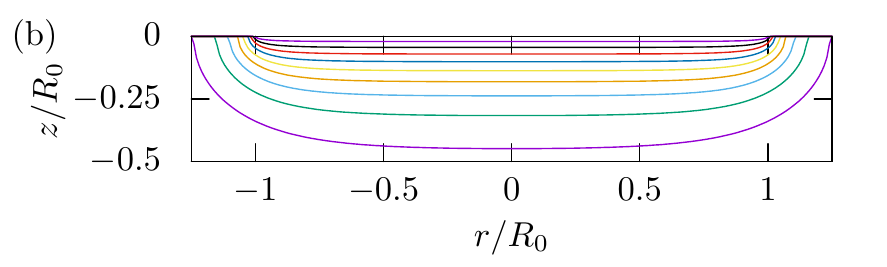}
        \end{minipage}
        \begin{minipage}[b]{0.45\linewidth}
        \includegraphics[width=7cm]{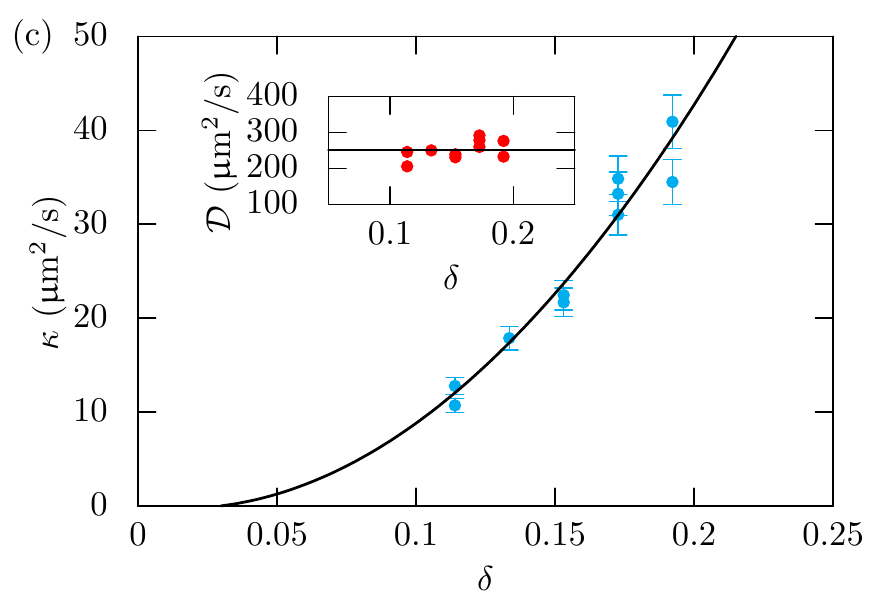}
        \end{minipage}
        \caption{
            Diffusion dynamics of water into a hydrogel.
            (a) Confocal image recorded during the absorption of a drop of fluorescent dye (CY5) in a gel ($\Gamma_a=0.175$) when two-third of the initial volume is absorbed by the gel. Several images acquired with a $10X$, $0.3$ N.A. objective are stitched to build the intensity image along the diameter. Iso-intensity profiles are superimposed on the picture. The scale bar represents $500$ $\mu$m.
            (b) Isoconcentration map of solvent concentration in the gel obtained from the solution of equation (\ref{eq:diffusion}).
            (c) Evolution of the effective diffusion coefficient $\kappa$ from equation (\ref{eq:flux_fit}) as a function of the swelling ratio $\delta$.
            The repeated experiments show the typical uncertainty and the solid line is a quadratic fit.
            The inset shows the resulting diffusion coefficient ${\cal D} = \kappa / (4 \beta^2 \delta^2)$.
            The average value is ${\cal D} = 250\pm 50$ $\mu$m$^2$/s.
        }\label{fig:absorption}
    \end{figure*}

    The final deposition pattern is the result of a combination of pinned and receding contact line regimes.
    In the following, we will analyze Regime I and describe theoretically the diffusion in the gel and the related absorption flux.
    We will show that during this regime, the absorption flux is homogeneous in the center of the drop and that an edge effect results in a larger flux increasing the particle concentration near the contact line.
    During regime II, because of the receding contact line, more liquid is absorbed in the drop center, erasing the larger concentration at the contact line.
At the end of the absorption process, we will show that the deposition is nearly homogeneous with a slight decrease of the concentration along the radius.

    \paragraph{Absorption flux}
    To quantify the gel's ability to absorb water, we define the swelling ratio $\delta$ as the difference of  water concentrations in the swollen and the initial states $\delta = \frac{V_{sw}^{w}}{V_{sw}^{tot}}  - \frac{V_i^{w}}{V_{i}^{tot}}$, where $V_i^{w}$ and $V_{sw}^{w}$ are, respectively, the volumes of water in the initial and swollen states, and $V_{i}^{tot}$ and $V_{sw}^{tot}$ are, respectively, the volumes of gel at the initial and swollen states.
    We checked that the swelling ratio increases linearly with the monomer concentration $\Gamma_a$.

    We measure the time evolution of the local adsorbed particle density to directly quantify the local flux of water from the drop to the gel.
    Images are taken in the plane of the gel surface with a fluorescent microscope automated with the software Micromanager \cite{Edelstein2010}.
Each sequence consists of 6 images in fluorescence followed by 1 image in bright field.
Sequences are repeated each 14 seconds until the end of the experiment.

The detection of the contact line on bright field images is done with Scikit-image by using a Canny filter and the circular Hough transform \cite{Vanderwalt2014}.
Thus, we obtain the radius and the center of the circle representing the position of the contact line on each picture.

We use the library Trackpy to perform particle tracking on images from fluorescent microscopy \cite{Allan2014}.
We analyze independently each sequence of 6 images.
This analysis consists first in the detection of the particle positions and then in the reconstruction of the particle trajectories.
From these particle trajectories, we extract those that have no significant displacement and correspond to the particles adsorbed on the gel interface.
To quantify the displacement, we use the standard deviation of the positions of the particles as follows.
For each trajectory consisting of a set of position $(r_i, \theta_i)$, we calculate the radial displacement $\delta r = \textrm{std}(r_i)$ and the orthoradial displacement $\bar{r} \delta \theta= \bar{r}\, \textrm{std}(\theta_i)$ where $\bar{r}$ is the average radial position and $\textrm{std}()$ is the standard deviation.
Thus, to identify adsorbed particles, we consider trajectories for which the radial displacement $\delta r < 4$ $\mu$m and the azimuthal displacement $\bar{r} \delta \theta < 4$ $\mu$m.
We proceed to a radial data binning consisting of 10 bins corresponding to identical gel surface areas.

    In Fig. \ref{fig:kinetics}b, we show the time evolution of the number of particles adsorbed at the surface of the gel in 10 concentric sectors of the same surface area $s = \pi R_0^{2}/10$.
    The outer sector, corresponding to the drop edge, has a typical width of $56$ $\mu$m.
    The other inner sectors are referred as the core and have a total surface area $9s$.
    With the exception of the drop edge, the time evolution of the number of absorbed particles is nearly independent of the radial position as shown  in Fig. \ref{fig:kinetics}b.
    Consequently, the accumulation of particles indicates that the flux $\bm{J}$ is independent of the radial position in the core region, \textit{i.e.} ${\bm J}(r,t) = J_c(t) {\bm e_z}$, where ${\bm e_z}$ is unit vector normal to the surface and $t$ is time.
     Thus, we define
       \begin{equation}
        J_c(t) =  \frac{2\,\pi}{9s} \int_{0}^{\sqrt{\frac{9}{10}} R_0}  J(r,t)\, r\mbox{d}r.
    \end{equation}

    Our results (Fig. \ref{fig:kinetics}b) show that the number of adsorbed particles evolves as $N = sC_p \sqrt{\kappa t}$ in each of the inner sectors, where $\kappa$ is a coefficient.
    From the number of adsorbed particles in the core region $N_{core}$, the flux of water (volume / unit area / unit time) to the gel in the core region is
    \begin{equation}
        J_c(t) =  \frac{1}{9s C_p} \frac{\textrm{d}N_{core}}{\textrm{d}t} = \frac{\sqrt{\kappa}}{2 \sqrt{t}}. \label{eq:flux_fit}
    \end{equation}

    \paragraph{Diffusion into the gel}

    To understand this time evolution, we consider experimentally and  theoretically the adsorption dynamics of a drop placed on a hydrogel.
Theoretically, a hydrogel can be modelled as a poroelastic material.
The swelling of a poroelastic material from Biot's theory applied to a polymer gel \cite{Doi2009,Yoon2010}
is composed of a mechanical equation for the displacements in the material, and a diffusion equation of the solvent in the material.
The boundary conditions generally involve the displacement and the chemical potential.
However, for weakly swelling gels, we neglect the effect of the displacement on the boundary conditions.
Thus, the set of equations can be reduced to a diffusion problem.
To model absorption during the Regime I, we consider a disk of radius $R_0$ diffusing in a semi-infinite medium described by a concentration $C(r,z,t)$.
    The semi-infinite medium assumption is supported by the observation reported in Fig. \ref{fig:absorption}a where the concentration gradient is established in the gel over a length-scale of approximatively $0.3$ mm, which is 10 times smaller than the gel thickness.
We introduce the reduced concentration $c(r,z,t)$ defined as $c=(C-C_{\infty})/(C_{sat}-C_{\infty})$ where $C_{\infty}$ is the initial concentration of water and $C_{sat}$ is the water concentration of a saturated gel.
The diffusion equation reads

   \begin{equation}
   \frac{\partial c}{\partial \tilde t} = \left( \frac{\partial^2 c}{\partial \tilde r^2} +  \frac{1}{\tilde r} \frac{\partial c}{\partial \tilde r} + \frac{\partial^2 c}{\partial \tilde z^2} \right),\label{eq:diffusion}
   \end{equation}
   where the dimensionless variables are $\tilde r =r/R_0$, $\tilde z = z/R_0$ and $\tilde t = t {\cal D}/R_0^2$.
This equation is supplemented by the boundary conditions: (i) $\partial_z c = 0$ for $z=0$ and $r>1$, (ii)
$c=1$ for $z=0$ and $r<1$ and (iii) $c=0$ everywhere at $t=0$.
On long timescales, the solution of this equation converges towards a stationary solution singular near $r=1$ \cite{Mandrik2001,Abdelrazaq2006}, but much
less is known on the time-dependent behaviour on short timescales, and on how the final field is reached.
Using singular perturbation techniques developed to solve
similar heat transfer problems \cite{Sneddon1966,Blackwell1972}, we were able to find an approximate analytical solution describing the short timescale behavior of the concentration field.
In this regime, the Laplace transform of the solution $\bar{c}  = \int_{0}^\infty c(r,z,t) e^{-\sigma t}\mbox{d}t$ reduces to

\begin{eqnarray}
\bar{c}
             \approx \frac{1}{2\sigma} \left(e^{-\sqrt{\sigma} \epsilon\eta} \mbox{erfc}\left\{ \frac{\sigma^{1/4} (\epsilon - \eta)}{\sqrt{2}} \right\}\right. \nonumber \\ +  \left. e^{\sqrt{\sigma} \epsilon\eta} \mbox{erfc}\left\{ \frac{\sigma^{1/4} (\epsilon + \eta)}{\sqrt{2}} \right\} \right)\label{solshort}
\end{eqnarray}
in which the following hyperbolic variables have been used

\begin{subequations}
    \begin{eqnarray}
        \epsilon &= \sqrt{-1+\frac{1}{4}\left(\sqrt{z^2+(r-1)^2}+\sqrt{z^2+(r+1)^2} \right)^2} \\
        \eta &= \sqrt{1-\frac{1}{4}\left(\sqrt{z^2+(r-1)^2}-\sqrt{z^2+(r+1)^2} \right)^2}.
    \end{eqnarray}
\end{subequations}
To obtain $c(r,z,t)$, equation (\ref{solshort}) is inverted numerically with Mathematica.

    To compare quantitatively the experimental observations with the theoretical predictions, the diffusion time $R_0^{2}/\cal{D}$ must be determined.
    We used the theoretical developments by Tanaka and Fillmore \cite{Tanaka1979} to measure independently this diffusion coefficient.
    Millimeter-diameter gel beads of the same composition are prepared with a millifluidic device composed of a T-junction by injecting the aqueous solution and a mineral oil.
    After crosslinking, the beads are placed in contact with water.
    The time evolution of the bead radii \cite{Tanaka1979} leads to a diffusion coefficient ${\cal D}=265\pm 40$ $\mu$m$^2$/s.

We performed experiments with a confocal microscope to quantify the concentration profiles in the gel.
    Specifically, a drop of water containing a fluorescent dye Sulfo-Cyanine5 azide (CY5, Lumiprobe, 4\% wt.) is absorbed on a gel and we observed the time evolution of the fluorescence signal in a vertical cross section along the drop diameter.
    Fig. \ref{fig:absorption}a shows one of these images at the end of Regime I.
    The fluorescence intensity beneath the drop in the core region is nearly flat and independent of the radial position whereas a rounded profile is observed at the drop edge.
    The solution of equation (\ref{eq:diffusion}) is presented in Fig. \ref{fig:absorption}b at the end of the Regime I, which corresponds to $\tilde t = 0.04$.
    As illustrated in Fig. \ref{fig:absorption}(a-b), we conclude that the fluorescent intensity distribution is in excellent agreement with the theoretical prediction, which validates the hypothesis of small deformations.
    We also observe that the gradient of water concentration in the gel weakly depends on the radial position in the core region.
    Finally, these results show that the distance over which the concentration varies in the gel is typically about 0.5 mm.
        This length-scale is smaller than the gel thickness and it validates the hypothesis of a semi-infinite material.


    \paragraph{Particle deposition}
    During Regime I in the core region, the time evolution of the average flux predicted by the solution of equation (\ref{eq:diffusion}) is given by
    \begin{equation}
        J_c = \beta {\cal D} \frac{\delta}{\sqrt{ {\cal D} t}}  = \beta \left( \frac{ {\cal D} \delta^2}{t} \right)^{1/2},\label{eq:flux_water_diffusion}
    \end{equation}
    where $\beta\approx 1.05$ is a numerical prefactor and $\delta$ characterizes the gel's swellability.
    The time dependence of the predicted flux is in agreement with our experimental measurements (Fig. \ref{fig:kinetics}b).

    \begin{figure}
        \centering
        \includegraphics[height=5cm]{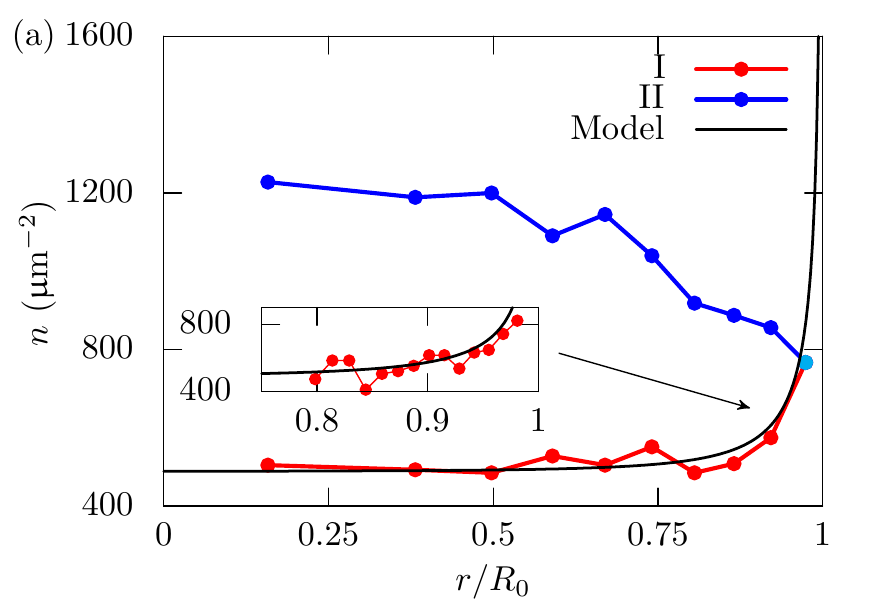}\\
        \includegraphics[height=4cm]{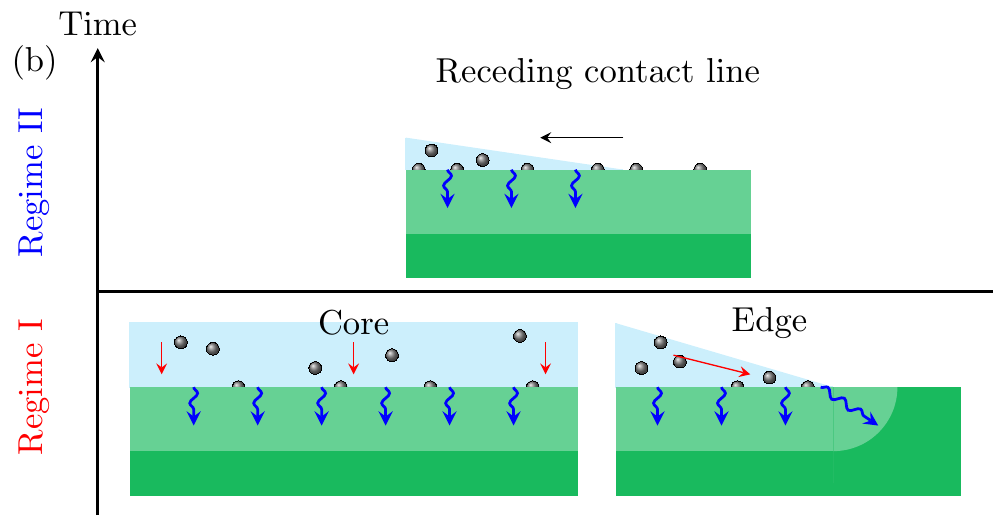}
        \caption{
        Explanation of the final deposition density profile.
            (a)  Particle density deposited on the surface along the radial position, at the end of Regime I (red) and Regime II (blue), respectively.
            The point in cyan indicates that the points corresponding to Regimes I and II are superimposed.
            The particle concentration is $C_p=4.0\times 10^{3}$ particles/mm$^3$.
            The solid black line represents the theoretical density of particles accumulated at the surface at the end of Regime I obtained from equation (\ref{eq:n_of_r}).
            The inset shows the radial increase of the particle density near the drop edge at the end of Regime I.
            (b) Schematics representing the particle deposition in Regimes I and II.
        }\label{fig:distribution}
    \end{figure}

    Combining equations (\ref{eq:flux_fit}) and (\ref{eq:flux_water_diffusion}), we obtain the diffusion coefficient ${\cal D} = \kappa / (4 \beta^2 \delta^2)$, as presented in the inset of Fig. \ref{fig:absorption}c.
    The range of swelling ratio $\delta$ investigated is limited. 
    For small $\delta$ values, the polymer chains at the gel surface are more hydrated and the wetting properties of the gel are changed. For large $\delta$ values, the weak swelling assumption for the theory could not be satisfied.
    By recording the number of particles accumulated versus time in the core region, we determine values of $\kappa$ for different initial swelling ratios $\delta$ of the hydrogels (Fig. \ref{fig:absorption}c).
    As shown in Fig. \ref{fig:absorption}c, the diffusion coefficient ${\cal D}$ is independent of the swelling ratio $\delta$.
    Consequently, the particle adsorption dynamics allows the measurement of the diffusion coefficient of water in the gel.
    The resulting value is ${\cal D} = 250 \pm 50$ $\mu$m$^2$/s is in good agreement with the measurement carried out by the swelling of gel beads.

    Furthermore, we observe in Fig. \ref{fig:kinetics}b a higher flux at the drop edge during Regime I.
    To highlight this effect, we plot in Fig. \ref{fig:distribution}a the radial particle distribution at the end of Regime I.
    The particle density deposited on the surface along the radial position, at the end of Regime I can be derived theoretically as
    \begin{equation}
        n(r) = C_p \int_0^{t_I} J(r,\tau) \mbox{d}\tau,\label{eq:n_of_r}
 \end{equation}
 where $t_I$ denotes the duration of Regime I and $J(r,\tau)$ is the flux of water from the drop to the gel at the radial position $r$.
    This profile is plotted in Fig. \ref{fig:distribution}a and shows a higher particle accumulation at the drop edge, which is quantitatively in excellent agreement with the measurements.

    During Regime II, the contact line recedes as illustrated in Fig. \ref{fig:kinetics}a.
    It is significant that the adhesion of the particles on the surface of the hydrogel is sufficiently high to prevent their transport by the receding contact line.
    Thus, the deposition occurs for a longer period in the drop center than at the edge.
    This effect counterbalances the edge effect that occurs during Regime I.
    For this reason, in the drop center, a larger particle number is observed in the final deposition pattern (Fig. \ref{fig:final}b and \ref{fig:distribution}).
    The difference on the final particle density between the center and the edge is less than $20$\% in the present experimental conditions.

We have demonstrated experimentally and theoretically that the absorption of a colloidal drop on an absorbing gel leads to a nearly uniform deposition pattern.
This uniformity is governed by the physical nature of the absorption phenomenon in swelling hydrogels, as we have quantified in equation (\ref{eq:flux_water_diffusion}).
We believe that the homogeneous coating of gels with particles can be used to tune locally their interfacial properties including roughness, wetting properties, rigidity, bonding capacities \cite{Rose2014} or permeability \cite{Dinsmore2002}.
Combined with simple microdispensing techniques such as ink-jet printing, our idea is amenable to produce scalable and complex graphics.
Such patterns offer the possibility to control the precise location and kinetics of drug delivery but also to inhibit microbial growth with silver particles that can be used on soft contact lenses or surgical implants \cite{Willcox2010,Gonzalez-Sanchez2015}.

    \paragraph{Acknowledgements}
    We thank H. Gelderblom, J. Nunes and M. Roch\'e for fruitful discussions.
    F.B. acknowledges that the research leading to these results received funding from the People Programme (Marie Curie Actions) of the European Union's Seventh Framework Programme (FP7/2007-2013) under REA grant agreement 623541.

    \paragraph{Author contributions}
    F.B. and F.I. contributed equally to this work.

    \bibliography{article}

\begin{thebibliography}{10}

\bibitem{Deegan1997}
R.D. Deegan, O.~Bakajin, T.F. Dupont, G.~Huber, S.R. Nagel, and T.A. Witten.
\newblock {\em Nature}, 389:827--829, 1997.

\bibitem{Deegan2000a}
R.~Deegan, O.~Bakajin, T.~Dupont, G.~Huber, S.~Nagel, and T.~Witten.
\newblock {\em Phys. Rev. E}, 62:756--765, 2000.

\bibitem{Marin2011}
A.G. Marin, H.~Gelderblom, D.~Lohse, and J.H. Snoeijer.
\newblock {\em Phys. Rev. Lett.}, 107:085502, 2011.

\bibitem{Kuang2014}
M.~Kuang, L.~Wang, and Y.~Song.
\newblock {\em Advanced Materials}, 26(40):6950--6958, 2014.

\bibitem{Kajiya2009}
T.~Kajiya, W.~Kobayashi, T.~Okuzono, and M.~Doi.
\newblock {\em J. Phys. Chem. B}, 113:15460--15466, 2009.

\bibitem{Still2012}
T.~Still, P.~Yunker, and A.~Yodh.
\newblock {\em Langmuir}, 28:4984--4988, 2012.

\bibitem{Sempels2013}
W.~Sempels, R.~De~Dier, H.~Mizuno, J.~Hofkens, and J.~Vermant.
\newblock {\em Nat Commun}, 4:1757, 2013.

\bibitem{Yunker2011}
P.~Yunker, T.~Still, M.~Lohr, and A.G. Yodh.
\newblock {\em Nature}, 476:308--311, 2011.

\bibitem{Wray2014}
A.~Wray, D.~Papageorgiou, R.~Craster, K.~Sefiane, and O.~Matar.
\newblock {\em Langmuir}, 30:5849--5858, 2014.

\bibitem{Biot1941}
M.~Biot.
\newblock {\em J. Appl. Phys.}, 12:155--164, 1941.

\bibitem{Aradian2000}
A.~Aradian, E.~Rapha{\"e}l, and P.G. de~Gennes.
\newblock {\em Eur. Phys. J. E}, 2:367--376, 2000.

\bibitem{Bacri2000}
L.~Bacri and F.~Brochard-Wyart.
\newblock {\em Eur. Phys. J. E}, 3:87--97, 2000.

\bibitem{Dou2012}
R.~Dou and B.~Derby.
\newblock {\em Langmuir}, 28:5331--5338, 2012.

\bibitem{Pack2015}
M.~Pack, H.~Hu, D.-O. Kim, X.~Yang, and Y.~Sun.
\newblock {\em Langmuir}, 0:null, 2015.

\bibitem{Doi2009}
M.~Doi.
\newblock {\em J. Phys. Soc. Jap.}, 78:052001, 2009.

\bibitem{Kajiya2011}
T.~Kajiya, A.~Daerr, T.~Narita, L.~Royon, F.~Lequeux, and L.~Limat.
\newblock {\em Soft Matter}, 7:11425--11432, 2011.

\bibitem{Hoare2008}
T.~Hoare and D.~Kohane.
\newblock {\em Polymer}, 49:1993--2007, 2008.

\bibitem{Hunt2014}
J.~Hunt, R.~Chen, T.~van Veen, and N.~Bryan.
\newblock {\em J. Mater. Chem. B}, 2:5319--5338, 2014.

\bibitem{Dinsmore2002}
A.D. Dinsmore, M.~Hsu, M.G. Nikolaides, M.~Marquez, A.R. Bausch, and D.A.
  Weitz.
\newblock {\em Science}, 298:1006--1009, 2002.

\bibitem{Benes2007}
K.~Benes, P.~Tong, and B.~Ackerson.
\newblock {\em Phys. Rev. E}, 76:056302, 2007.

\bibitem{GomezSuarez1999}
C.~G\'omez~Su\'arez, J.~Noordmans, H.~C. van~der Mei, and H.~J. Busscher.
\newblock {\em Langmuir}, 15(15):5123--5127, 1999.

\bibitem{Sudre2011}
G.~Sudre.
\newblock {\em {T}unable {A}dhesion of {H}ydrogels}.
\newblock PhD thesis, Universit{\'e} Pierre et Marie Curie - Paris VI, 2011.

\bibitem{Edelstein2010}
A.~Edelstein, N.~Amodaj, K.~Hoover, R.~Vale, and N.~Stuurman.
\newblock {\em {C}omputer {C}ontrol of {M}icroscopes {U}sing {M}icromanager}.
\newblock John Wiley \& Sons, Inc., 2010.

\bibitem{Vanderwalt2014}
S.~van~der Walt, J.L. Sch{\"o}nberger, J.~Nunez-Iglesias, F.~Boulogne, J.D.
  Warner, N.~Yager, E.~Gouillart, and T.~Yu.
\newblock {\em PeerJ}, 2:e453, 2014.

\bibitem{Allan2014}
D.~Allan, T.~Caswell, and N.~Keim.
\newblock 2014.

\bibitem{Yoon2010}
J.~Yoon, S.~Cai, Z.~Suo, and R.~Hayward.
\newblock {\em Soft Matter}, 6:6004--6012, 2010.

\bibitem{Mandrik2001}
P.A. Mandrik.
\newblock {\em Mathematical Modelling and Analysis}, 6:280--288, 2001.

\bibitem{Abdelrazaq2006}
N.~Abdelrazaq.
\newblock {\em J. Math. Stat.}, 2:346, 2006.

\bibitem{Sneddon1966}
I.~Sneddon.
\newblock {\em {M}ixed {B}oundary {V}alue {P}roblems in {P}otential {T}heory}.
\newblock North Holland Publ. Amsterdam, 1966.

\bibitem{Blackwell1972}
J.H. Blackwell.
\newblock {\em Journal of the Australian Mathematical Society}, 14:433--442,
  1972.

\bibitem{Tanaka1979}
T.~Tanaka and D.~Fillmore.
\newblock {\em J. Chem. Phys.}, 70:1214--1218, 1979.

\bibitem{Rose2014}
S.~Rose, A.~Prevoteau, P.~Elziere, D.~Hourdet, A.~Marcellan, and L.~Leibler.
\newblock {\em Nature}, 505:382--385, 2014.

\bibitem{Willcox2010}
M.P. Willcox, E.H. Hume, A.~Vijay, and R.~Petcavich.
\newblock {\em Journal of Optometry}, 3:143--148, 2010.

\bibitem{Gonzalez-Sanchez2015}
M.I Gonz{\'a}lez-S{\'a}nchez, S.~Perni, G.~Tommasi, N.~Morris, K.~Hawkins,
  E.~L{\'o}pez-Cabarcos, and P.~Prokopovich.
\newblock {\em Materials Science and Engineering: C}, 50:332--340, 2015.

\end{thebibliography}
    \bibliographystyle{unsrt}

    \end{document}